\documentclass[review]{elsarticle}
\usepackage{lineno}

\usepackage{multirow,setspace, amssymb, amsmath, graphicx, color, rotating, subfigure}
\usepackage{lineno}
\usepackage{textcomp}
\usepackage{booktabs}    
\usepackage{CJK}
\usepackage{bm}%
\usepackage{rotating}
\usepackage{diagbox}
\usepackage{epsfig}
\usepackage{dcolumn}
\usepackage{epstopdf}
\usepackage{natbib}
\biboptions{sort&compress}
\usepackage[hidelinks]{hyperref}

\journal{*}

\begin{document}

\begin{frontmatter}

\title{Directed network comparison using motifs}

\author[inst1]{Chenwei Xie}
\author[inst1]{Qiao Ke}
\author[inst1]{Haoyu Chen}
\author[inst1]{Chuang Liu}
\author[inst1,inst2]{Xiu-Xiu Zhan\corref{ca1}}\ead{zhanxiuxiu@hznu.edu.cn}

\address[inst1]{Alibaba Research Center for Complexity Sciences, Hangzhou Normal University, Hangzhou, 311121, P. R. China}
\address[inst2]{College of Media and International Culture, Zhejiang University, Hangzhou 310058, PR China}

\cortext[ca1]{Corresponding authors.}

\begin{abstract}
Analyzing and characterizing the differences between networks is a fundamental and challenging problem in network science.  Previously, most network comparison methods that rely on topological properties have been restricted to measuring differences between two undirected networks. However, many networks, such as biological networks, social networks, and transportation networks, exhibit inherent directionality and higher-order attributes that should not be ignored when comparing networks.  Therefore, we propose a motif-based directed network comparison method that captures the local, global, and higher-order differences between two directed networks. Specifically, we first construct a motif distribution vector for each node, which captures the information of a node's involvement in different directed motifs. Then, the dissimilarity between two directed networks is defined on the basis of a matrix which is composed of the motif distribution vector of every node and Jensen-Shannon divergence. The performance of our method is evaluated via the comparison of six real directed networks with their null models as well as their perturbed networks based on edge perturbation. Our method is superior to the state-of-the-art baselines and is robust with different parameter settings.

\end{abstract}


\end{frontmatter}

\section{Introduction}
\label{Intro}
\makeatletter
\newcommand{\rmnum}[1]{\romannumeral #1}
\newcommand{\Rmnum}[1]{\expandafter\@slowromancap\romannumeral #1@}
\makeatother

Many systems in various domains, featuring intricate interaction relationships, can be effectively represented in the form of complex networks~\cite{barabasi2013network}, including social platforms~\cite{knoke2019social, zhan2020susceptible}, biological systems~\cite{liu2020computational}, economic systems~\cite{schweitzer2009economic}.
Due to the diversity of network forms~\cite{bretto2013hypergraph, kivela2014multilayer} and the high-order features of networks~\cite{benson2016higher, xie2023vital}, the precise measurement of similarity between different networks, namely the design of an effective network comparison method, has emerged as a central focus in the field of network science. 
Network comparison aims to quantify the differences between two networks based on network topological structure, allowing the effective handling of different types of tasks~\cite{tantardini2019comparing, soundarajan2014guide}. For example, in the field of pattern recognition, network comparison can be applied to classify content such as images, documents, and videos~\cite{conte2004thirty}. In the biological domain, network comparison can be used to analyze which protein interactions may have equivalent functions~\cite{sharan2006modeling}. In neuroscience, the comparison of brain networks contributes to understanding the functional differences between normal and pathological brains~\cite{mheich2020brain}.

The original term used to compare networks was the graph isomorphism problem~\cite{zemlyachenko1985graph}, which has been proven to fall within the NP complexity class~\cite{cook2023complexity}.
In recent years, researchers have proposed various methodologies from different perspectives and technologies to measure the similarity between networks~\cite{latora2007measure, xiao2008symmetry, babai2016graph, lv2019eigenvector,wang2020path, schieber2017quantification}. The majority of these methods have primarily concentrated on the comparison of undirected networks. However, interactions among distinct entities in the real world commonly exhibit asymmetry. In social networks, an instance of user $i$ trusting user $j$ does not necessarily imply reciprocal trust from $j$ to $i$. The directionality of the interactions between nodes in a network, which cannot be captured by an undirected network, has boosted the research of directed network comparison. For example, Bagrow and Bollt~\cite{bagrow2019information} utilized portrait divergence, a metric based on the distribution of the shortest path lengths, to evaluate the structural similarities between networks. Koutra et al.~\cite{koutra2013deltacon} proposed DeltaCon by calculating the Matusita distance of similarity matrices between two networks. Sarajlic et al.~\cite{sarajlic2016graphlet} extended network distance measures to directed networks using directed graphlets, demonstrating their efficacy in distinguishing various directed networks. The centrality-based methods, such as degree~\cite{prvzulj2007biological}, closeness~\cite{cohen2013scalable}, and clustering coefficient~\cite{yaverouglu2015proper}, compare networks based on the centrality values of each node. Although these methods are capable of comparing networks effectively to some extent, most of them have not considered the higher-order structure of a network, which has been shown to be ubiquitous in various complex systems~\cite{xie2023vital}. Consequently, we propose using direct motifs to quantify the dissimilarity between two networks. Motifs refer to recurring subgraphs in a network, where these subgraphs exhibit specific interaction patterns that facilitate understanding of the functionality of networks~\cite{milo2002network}. Motifs have been widely used in different network tasks, i.e., community detection\cite{li2018community}, link prediction\cite{qiu2021temporal}, and node ranking problems\cite{zhao2023novel}.In contrast to traditional conventional methods, motif-based approaches consistently exhibit superior performance in tackling these problems.

To explore the similarity between different directed network structures, in this paper we propose a motif-based directed network comparison method $D_m$, i.e., using motifs to examine smaller components of directed networks to assess the similarity between networks. We start by constructing a node motif distribution matrix, where the elements in the matrix are obtained by computing the distribution of nodes appearing in different directed motifs. Due to computational complexity, we consider the motifs composed of $2$ to $4$ nodes and thus obtain $35$ different directed motifs. Later on, we use the Jensen-Shannon divergence to quantify the dissimilarity between two directed networks both locally and globally. We validate the effectiveness of $D_m$ in six real directed networks. Compared to the baseline methods, $D_m$ exhibits notable distinguishability and robustness in comparing networks.

The rest of the paper is organized as follows. Section \ref{2} introduces the definition of motifs in a directed network and details the motif-based directed network comparison method. We provide a clear description of the baseline methods and directed network datasets in Section \ref{3}. All experiment results are presented in Section \ref{4}. Section \ref{5} summarizes the full paper.

\section{Method}
\label{2}
\subsection{The definition of motifs in a directed network}
A directed unweighted network is represented as $G = (V, E)$, where $V=\left \{v_1,v_2,\cdots, v_N  \right \}$ and $E=\left \{ e_k =\left ( v_{i},v_{j}   \right )|k=1,\cdots, M|v_{i},v_{j}\in V  \right \}$
are the node and edge set, respectively. The number of nodes and the number of edges are given by $N$ and $M$. The adjacent relationship between two nodes in $G$ is given by the adjacency matrix $A$, with $A_{ij} = 1$ indicating that there is a directed edge between $v_i$ and $v_j$, and $A_{ij} = 0$ implying that there are no edges between them. We note that the directionality of $G$ determines that $A$ is an asymmetric matrix.

\begin{figure}[htpb]
    \centering
    \includegraphics[width=1\linewidth]{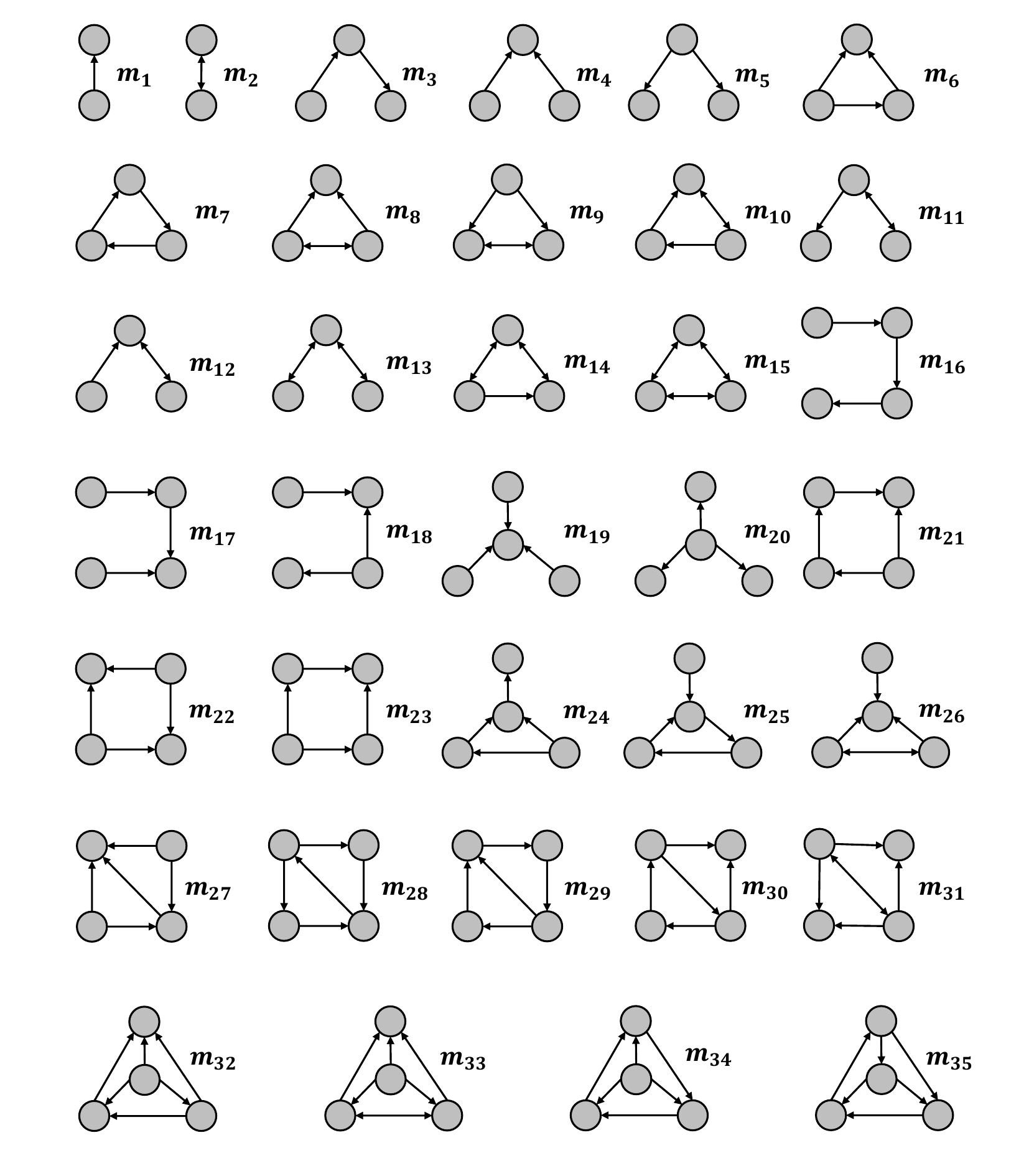}
    \caption{Motifs formed by 2 to 4 nodes in directed networks. All the motifs are labeled from $m_1$ to $m_{35}$.}
    \label{fig:motif}
\end{figure}

Motifs are the most common graphical patterns in complex networks, consisting of a group of closely connected nodes and edges. Due to the high complexity of computing motifs in a network, we normally consider motifs formed by $2$ to $4$ nodes.
 Motifs play a crucial role in the study of complex networks, acting as fundamental building blocks for large complex networks, analogous to genes in biology. In a directed network, the motifs are formed by nodes with directed edges. We show examples of directed motifs in Figure~\ref{fig:motif}. There are $35$ directed motifs, each comprising $2$ to $4$ nodes, individually represented as $m_1$ to $m_{35}$, respectively. For instance, there are two kinds of motifs if we consider two nodes, which are given by $m_1$ and $m_2$ in the figure.

\subsection{Motif-Based Directed Network Comparison Method}
Motifs contain important topological information of a network and thus are essential for network comparison. Based on the distinctive topological properties of directed motifs, we first compute the motif distribution in a directed network. As the time complexity of computing motifs is quite high, we will use the motifs listed in Figure~\ref{fig:motif} that are formed by $2,3$, and $4$ nodes for the computation of motif distribution.
Specifically, we use $T_{i} =\left \{ t_i(j)|1\le j\le 35 \right \}$ to represent the motif distribution of node $v_i$, where $t_i(j)$ represents the fraction of motif $j$ that contains $v_i$. Consequently, a $N \times 35$ matrix $\mathcal{T} = \left \{ T_{1}, T_{2}, \cdots, T_{N} \right \}$ can be constructed based on the motif distribution of every node. We further define directed network node dispersion ($DNND$) to measure connectivity heterogeneity between nodes. A larger $DNND$ indicates greater heterogeneity in the connectivity of nodes within the network, while a smaller $DNND$ suggests a more uniform distribution of node connections. And $DNND$ is given by the following equation:
\begin{equation}
    DNND(G)=\frac{\zeta(T_{1}, T_{2}, \cdots, T_{N}) }{\ln_{}{(N+1)} }, 
\end{equation}
where $\zeta(T_{1}, T_{2}, \cdots, T_{N})$ is the Jensen-Shannon divergence of the $N$ motif distributions, and is given by:
\begin{equation}
    \zeta(T_1, T_2, \cdots, T_N)=\frac{1}{N} \sum_{i,j}^{}t_i(j)\ln_{}{(\frac{t_i(j)}{\mu _j} )},  
\end{equation}
where $\mu_j$ represents the average value of $N$ motif distributions, the specific calculation is as follows:
\begin{equation}
    \mu _j=\frac{\sum_{i=1}^{N}t_i(j)}{N}
\end{equation}

Given two directed networks $G_1\left ( V_1, E_1 \right ) $ and $G_2\left ( V_2, E_2 \right )$, the structural dissimilarity between them can be calculated based on their motif distribution matrices $\mathcal{T}_1$ and $\mathcal{T}_{2}$. We use $D_m(G_1, G_2)$ to represent the dissimilarity between $G_1$ and $G_2$, thus,
 
\begin{equation}
    D_m(G_1,G_2)=\varphi \sqrt{\frac{\zeta (\mu^{G_1},\mu^{G_2} )}{\ln_{}{2} } }+(1-\varphi)\left | \sqrt{DNND(G_1)}-\sqrt{DNND(G_2)}   \right |
\end{equation}

The dissimilarity $D_m$ comprises two terms,  and we use a parameter $\varphi (0 \leq \varphi \leq 1)$ to adjust their weights. The first term illustrates the difference between the average motif distributions, that is, $\mu^{G_1}=(\mu^{G_1}_1, \mu^{G_1}_2, \cdots, \mu^{G_1}_{N_1})$ and $\mu^{G_2}(\mu^{G_2}_1, \mu^{G_2}_2, \cdots, \mu^{G_2}_{N_2})$, and predominantly signifies the global distinctions between the two networks. The second term mainly describes the difference between the $DNND$s of the two networks, indicating the local difference between them. A lower value of $D_m$ indicates a higher network similarity and vice versa.

\section{Baselines and Datasets}\label{3}
\subsection{Baselines}
\textbf{Portrait-based directed network comparison method~\cite{bagrow2019information}:} 
For a directed network $G$, we construct a portrait matrix $B$ based on the distance between nodes. Each element $B_{l,k}$ represents the number of nodes that have $k$ nodes at distance $l$, where $0 \leq l \leq d$, $0 \leq k \leq N-1$, and $d$ represents the diameter of $G$. We note that we utilize the shortest directed path length to calculate the distance between nodes. In addition, $B$ is independent of the ordering and labeling of the nodes. Based on $B_{l,k}$, we can derive the probability that a randomly selected node has $k$ nodes at a distance of $l$ and is given by

\begin{equation}
    Q_{l,k}=\frac{1}{N} B_{l,k} 
\end{equation}

For two directed networks, $G_1$ and $G_2$, the probability distributions $Q_1$ and $Q_2$ are employed to interpret the rows of the network portraits for each of them. The similarity between $G_1$ and $G_2$ is represented as $D_p\left ( G_1,G_2 \right ) $ and is defined as:
\begin{equation}
    D_p\left ( G_1,G_2 \right )=\frac{1}{2}KL\left ( Q_1||M \right )+\frac{1}{2}KL\left ( Q_2||M \right ),
\end{equation}
where $M=\frac{1}{2}\left ( Q_1+Q_2 \right ) $, $KL\left ( *||* \right )$ represents the Kullback-Liebler divergence between two distributions.

\textbf{DeltaCon-based directed network comparison method~\cite{koutra2013deltacon}:} DeltaCon considers the similarity between two networks by quantifying the difference of the r-step paths other than the edges. Given a directed and unweighted network $G$ and its adjacency matrix $A$, the r-step paths are encoded in the similarity matrix $S=\left [ I+\varepsilon ^{2}D-\varepsilon A  \right ]^{-1}$, where
$D$ and $I$ are diagonal matrices with diagonal elements equal to node degree and $1$, respectively, and $\varepsilon = 1/(1+\max(D_{ii})) (i=1,\cdots, N)$. We assume that the similarity matrices for two directed and unweighted networks $G_1$ and $G_2$ are denoted as $S$ and $S^{'}$, and the dissimilarity $D_d$ between them is given by the following equation:

\begin{equation}
    D_d\left ( G_1,G_2 \right )=\left \{ \sum_{i,j=1}^{N}\left ( \sqrt{S_{ij}}-\sqrt{S^{'}_{ij}}   \right )^2   \right \}^{\frac{1}{2}}  
\end{equation}

\textbf{Closeness-based directed network comparison method:}  Centrality measures, such as degree, betweenness, and closeness, were used to compare networks~{\cite{cohen2013scalable}}. However, in the part of experiments, we find that closeness centrality surpasses other centrality methods in network comparison. Therefore, we omit the other centrality measures and only use closeness for directed network comparison.
Closeness centrality measures the importance of a node within a network by evaluating the proximity of its connections to other nodes. The closeness centrality of a node is defined as

\begin{equation}
    c_i=\frac{1}{ {\textstyle \sum_{i\ne j}^{}d_{ij}} },  
\end{equation}
where $d_{ij}$ represents the directed shortest path length from node $v_i$ to node $v_j$. For two directed networks $G_1$ and $G_2$, we assume that the closeness centrality vectors for them are given by $c=\left ( c_1, c_2, \cdots, c_N \right ) ^T$ and $c^{'}=\left ( c_1^{'}, c_2^{'}, \cdots, c_N^{'} \right ) ^T$. 
Therefore, the dissimilarity between $G_1$ and $G_2$ based on closeness centrality is given as follows:
\begin{equation}
    D_{c}\left ( G_1,G_2 \right )=\sum_{i=1}^{N}\left | c_i-c'_i \right |   
\end{equation}

\subsection{Description of  Directed Network Datasets}
To evaluate the performance of our proposed methods and the state-of-the-art baselines, we select six real-world directed networks from diverse domains including biological networks, transportation networks, and social networks. The descriptions of each of the datasets are as follows:

\textbf{Mac\cite{takahata1991diachronic}} describes the interactions between adult female Japanese macaques, and is about the dominance behavior between them. Each node denotes a macaque and a directed edge from node $v_i$ to $v_j$ indicates the dominance of $v_i$ over $v_j$.

\textbf{Caenorhabditis elegans (Elegans)\cite{white1986structure}} is a neural network of Caenorhabditis elegans. It uses directed edges to represent neural connections among neurons in the nervous system of Caenorhabditis elegans.

\textbf{Physicians\cite{coleman1957diffusion}} is a directed network that describes the spread of innovation among physicians. A directed edge $(v_i, v_j)$ between two physicians $v_i$ and $v_j$ implies that $v_i$ would turn to $v_j$ if he needs suggestions or is interested in a discussion.

\textbf{Email-Eu-core (Email)\cite{leskovec2007graph}} is an email network that captures email interactions between institution members in a large European research institution. A directed edge between two staff $v_i$ and $v_j$ means that staff $v_i$ has sent an email to staff $v_j$.

\textbf{US airport\cite{kunegis2013konect}} illustrates the flight connections between US airports. A directed edge $(v_i, v_j)$ between two airports $v_i$ and $v_j$ illustrates that there is at least a flight from airport $v_i$ to $v_j$.

\textbf{Chess\cite{kunegis2013konect}} is a network that characterizes the interaction between players in an international chess game within a month. A directed edge is formed from a white player to a black player in this network.

Table~\ref{tab1} shows the basic properties of the directed networks mentioned above, including the number of nodes $\left ( N \right ) $, the number of edges $\left ( M \right ) $, average degree $\left ( Ad \right ) $, average shortest path length $\left ( Avl \right ) $, and network diameter $\left (d\right ) $.

\begin{table}[htpb]
\centering
\caption{Basic properties of real directed networks, where $N $, $M$, $Ad$, $Avl$, and $d$ represent the number of nodes, the number of edges, average degree, average shortest path length, and network diameter, respectively.\label{tab1}}

\begin{tabular}{cccccc}
\hline
\textbf{Networks} & \textbf{$N$} & \textbf{$M$} & \textbf{$Ad$} & {$Avl$} & \textbf{$d$}\\ \hline
Mac & 62			& 1187  &38.29  &1.38   &2\\
Elegans   & 237   & 4296  &28.92  &2.47   &5 \\
Physicians     &241    &1098   &9.11   &2.58   &4\\
Email       &1005   &25571  &50.84  &2.94   &7\\
US airport    &1574   &28236  &35.87  &3.13   &8\\
Chess    &7301   &65053  &17.82  &3.92   &13\\\hline
\end{tabular}
\end{table}

\section{Experimental Results}
\label{4}
\subsection{The dissimilarity between a real network and its null models}
The null model is widely used as a tool for the comparison of network topology\cite{wang2022quantification}, which retains specific network properties, such as degree distribution or clustering coefficient via random reshuffling of network connections. In this section, we propose three null models for directed networks to gradually change the network topology and use our comparison method to compare each directed network and its null models.

We extend the $dk$-series null models that were originally proposed for undirected networks to directed networks\cite{orsini2015quantifying}, which retain the degree distributions, correlations, and clustering of a real directed network to some extent. Concretely, the models are illustrated as follows:
$Dk1.0$ preserves the outdegree and indegree of a node by randomly rewiring each directed edge. Therefore, the degree sequence of the original network is preserved in the reshuffling process.
$Dk2.0$ reshuffles every edge in the network while maintaining the outdegree, indegree, and joint degree distribution of the original network.
$Dk2.5$ rewires every edge by preserving the distribution of the degree-dependent clustering coefficient. 
We note that the newly formed directed edges should never have existed in the original network before.

\begin{figure}[h]
    \centering
    \includegraphics[width=1\linewidth]{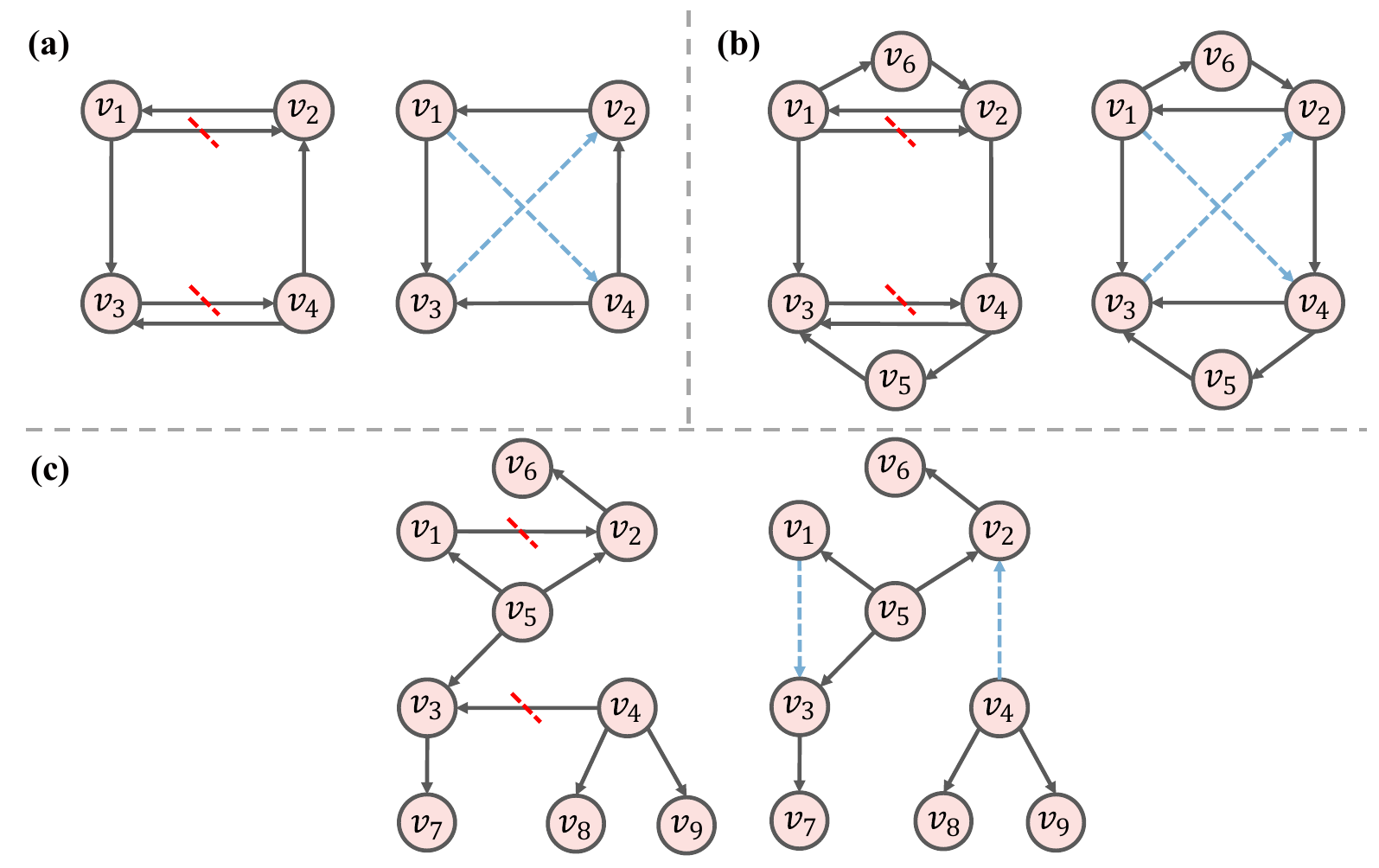}
    \caption{Toy examples of three $dk$-series null models: (a)$Dk1.0$; (b)$Dk2.0$; (c)$Dk2.5$. The blue dashed lines indicate the newly connected edges. In (a), (b), and (c), the left panel shows the original network and the right panel shows the rewired network. }
    \label{fig:null_model}
\end{figure}

We show examples of how to generate the null models in Figure \ref{fig:null_model}(a-c), the blue dashed lines indicate the newly connected edges.  The left panel shows the original network, and the right panel shows the network after the rewiring process in each of the figures. Figure \ref{fig:null_model}(a) shows an instance for $Dk1.0$. Specifically, we disconnect the edges $\left ( v_{1},v_{2} \right )$ and $\left ( v_{3},v_{4} \right )$, and form new edges, i.e., $(v_{1}$, $v_{4})$ and $(v_{3}$, $v_{2})$. Therefore, the in-degree and out-degree of each node is preserved in this process. Figure \ref{fig:null_model}(b) demonstrates the generation of a random network via $Dk2.0$, which is more strict than $Dk1.0$. For example, if we disconnect the directed edge between $v_1$ and $v_2$, that is, $(v_1, v_2)$, we need to find a node that has the same indegree and outdegree as $v_2$, and the appropriate node is $v_4$. Accordingly, we connect $v_1$ and $v_4$ and form a new directed edge $(v_1,v_4)$. Therefore, $Dk2.0$ maintains the degree sequence and the joint degree distribution of a network. In Figure \ref{fig:null_model}(c), the degree (sum of indegree and outdegree) and clustering coefficient for each node are $\{2, 3, 3, 3, 3, 1, 1, 1, 1\}$ and $\{1/2, 1/6, 0, 0, 1/6, 0, 0, 0, 0\}$, respectively. Therefore, the average clustering coefficients for nodes that have degree of $\{1, 2, 3\}$ are $\{0, 1/2, 1/12\}$, respectively, which are also called degree-dependent clustering coefficients. We disconnect the directed edges $(v_1,v_2)$ and $(v_4,v_3)$ and form new directed edges as $(v_1,v_3)$ and $(v_4, v_2)$. In the rewired network, the degree-dependent clustering coefficient distribution is the same as the original network.

\begin{figure}[h]
    \centering
    \includegraphics[width=0.50\linewidth]{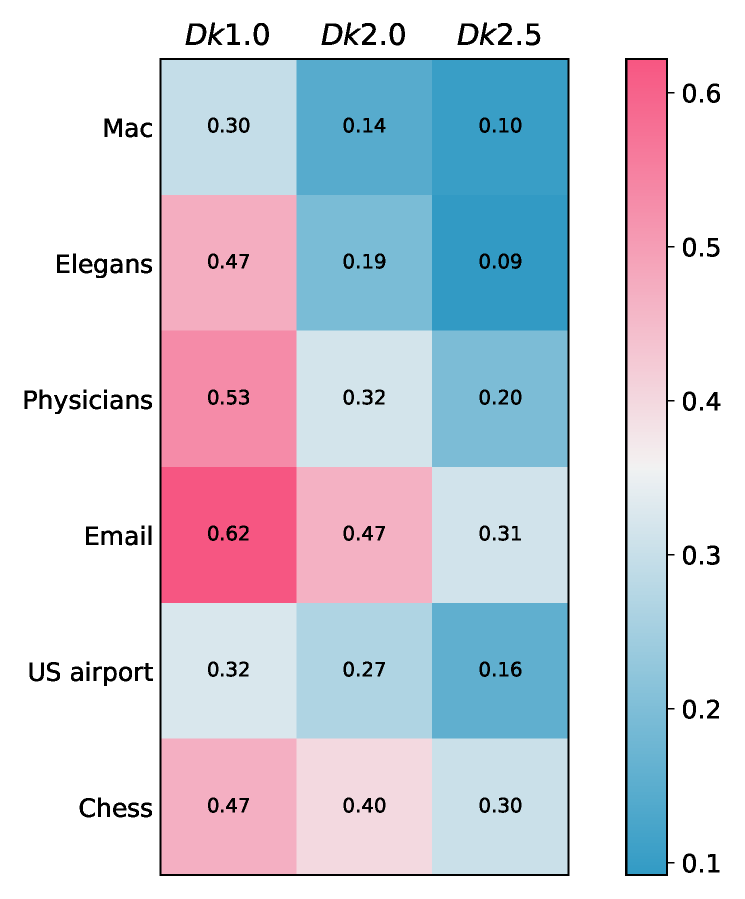}
    \caption{Comparison between real directed networks and their null models via motif-based directed network comparison method. The null models are $Dk1.0$, $Dk2.0$ and $Dk2.5$. Smaller values in the heatmap indicate a higher similarity, and vice versa. }
    \label{fig:nullmodel}
\end{figure}

A lower value of $k$ implies greater disruption of the original network structure.
In Figure~\ref{fig:nullmodel}, we use the motif-based directed network comparison method to quantify the dissimilarity between each of the directed networks and its three null models. Experimental results across six networks suggest that as $k$ increases, the similarity between the original network and its null models gradually increases. The dissimilarity observed in our approach aligns with the generation of null models, providing further confirmation of the effectiveness and stability of our model for comparing directed networks from different domains.

\subsection{The comparison of directed network and its perturbed network}
\begin{figure}[h]
    \centering
    \includegraphics[width=1\linewidth]{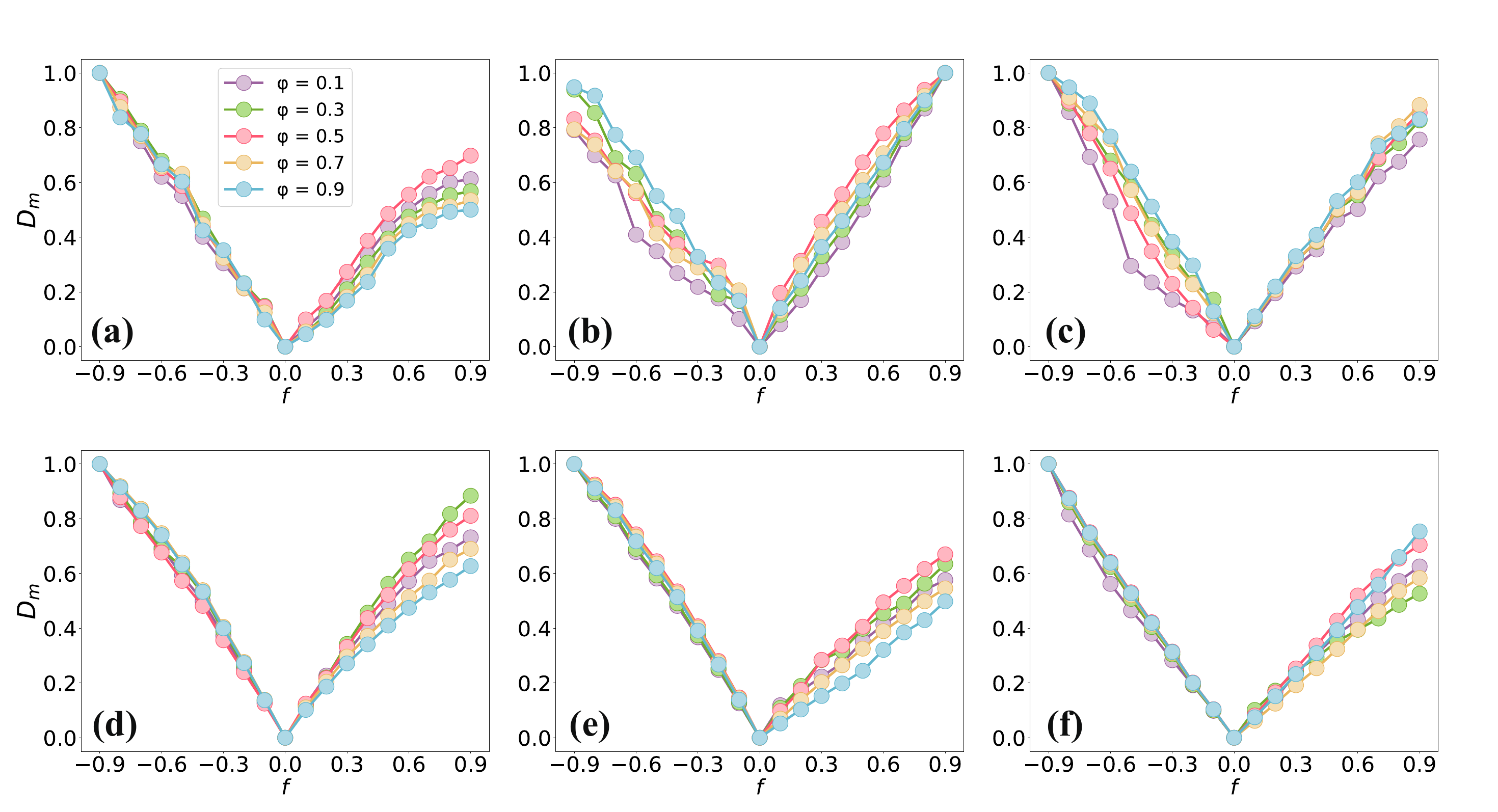}
    \caption{Similarity between a real directed network and perturbed network generated by randomly adding or deleting edges, where positive values of $f$ indicate we randomly add $f$ fraction of edges, and vice versa. We show results for networks: (a) Mac; (b) Elegans; (c) Physicians; (d) Email; (e) US airport; (f) Chess. The parameter $\varphi$ of $D_m$ is set to $0.5$. Each point in the figure is averaged over 100 realizations.}
    \label{fig:edge-add-dele}
\end{figure}

In this section, we perform perturbation experiments on the edges of six real directed networks to further assess the stability and applicability of the motif-based comparison method. Specifically, for each given network, we randomly add or remove edges with a certain proportion $f$, where the range of $f$ is $\left [-0.9, 0.9 \right ]$. The positive value of $f$ indicates that we randomly add $|f|$ fraction of directed edges into the network, and the negative value of $f$ means that we randomly remove $|f|$ fraction of the directed edges. We compare the original network with the perturbed network by adding or removing edges using different network comparison methods, as shown in Figure~\ref{fig:edge-add-dele}. The four comparison methods ($D_m$, $D_p$, $D_d$, and $D_c$) show similar trends; that is, the increase in $|f|$ will make the perturbed network have a greater difference from the original network, which is consistent with intuition. This conclusion is especially significant when $f$ is negative. However, the motif-based comparison method is much better than the rest of the baselines for positive values of $f$. The curves of the other three baselines for $f>0$ are flatter than those of our method. Taking the Mac network as an example (Figure~\ref{fig:edge-add-dele}(a)), the values of $D_p$ range from $0.07$ to $0.13$ for $f \in [0,1]$, and the values of $D_p$ are the same for $f=0.1$ and$f=0.2$, which is unreasonable. $D_d$ and $D_c$ also show insignificant dissimilarities between networks in Figure~\ref{fig:edge-add-dele}(a)-(f). The baseline methods, such as $D_p$ and $D_c$, are based on the distance between nodes, and $D_d$ considers the r-step paths of a network for network comparison. However, they have not considered the higher-order network structure of a network and thus may result in poor performance in network comparison.

\subsection{Parameter Sensitivity Analysis}
\begin{figure}[h]
    \centering
    \includegraphics[width=1\linewidth]{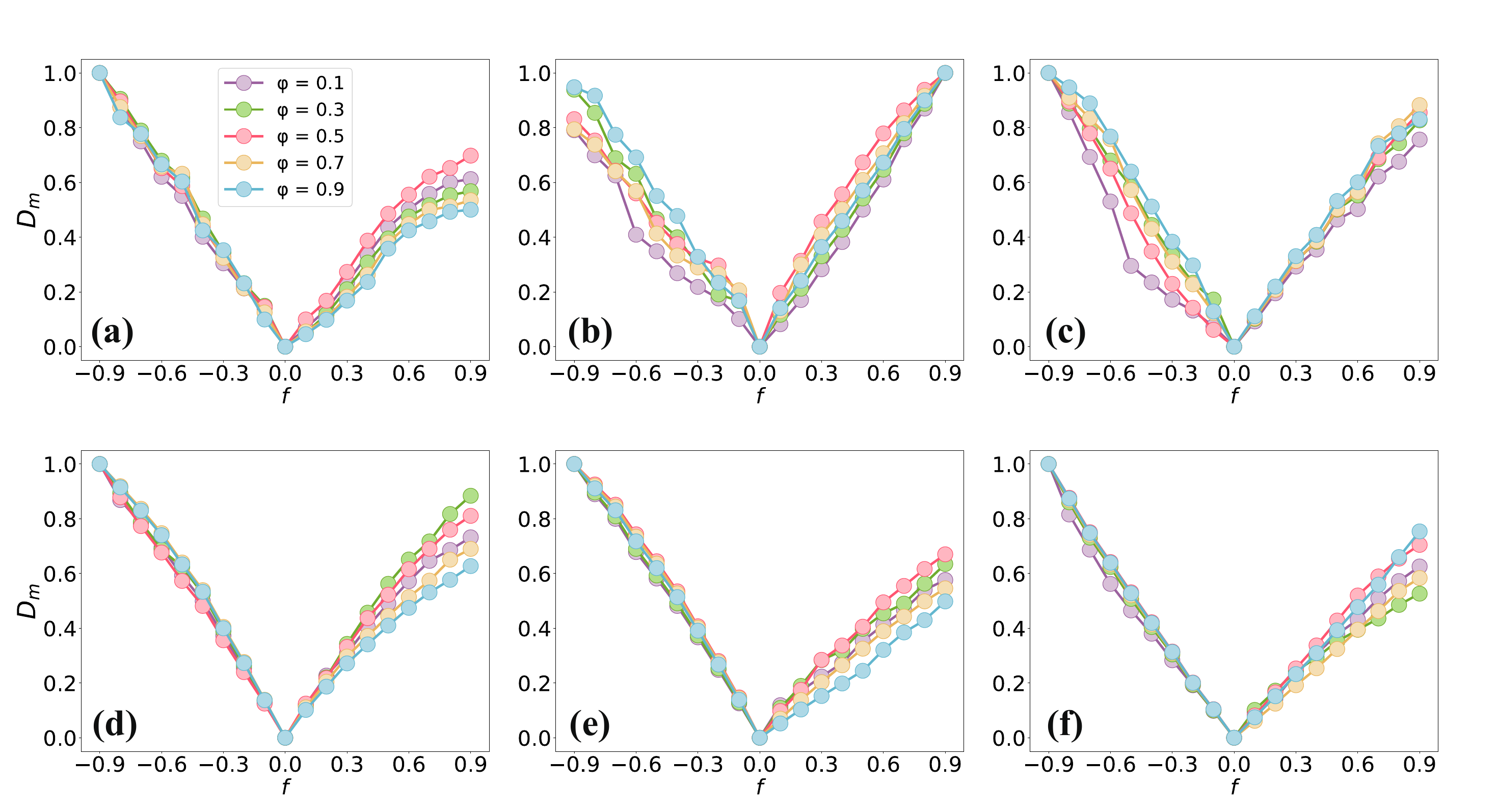}
    \caption{Parameter analysis for motif-based directed network comparison. We compare the real network with its perturbed network via edge addition or deletion. Different curves show we choose different values of  $\varphi$, which is the only parameter in our method, $\varphi \in \left \{0.1, 0.3, 0.5, 0.7, 0.9\right \}$. Positive values of $f$ indicate the random edge addition, and vice versa. We show results for networks: (a) Mac; (b) Elegans; (c) Physicians; (d) Email; (e) US airport; (f) Chess. All results are averaged over 100 realizations.}
    \label{fig:para-analy}
\end{figure}

The motif-based directed network comparison method involves a parameter, denoted as $\varphi$, that determines how much importance is given to the global or local differences between two networks, with larger value of $\varphi$ indicting we consider more of global difference and vice versa. Therefore, we perform parameter analysis for $\varphi$ in the six real-world directed networks via the comparison of original network and its perturbed networks. The results are given in Figure~\ref{fig:para-analy}, in which we use curves with different colors indicating we choose different values of $\varphi (\varphi \in \{0.1, 0.3, 0.5, 0.7, 0.9\})$. The figure displays curves that exhibit a similar trend for different values of $\varphi$, and there is small deviation among the curves when $f<0$. However, the network dissimilarity for different $f$ is more significant for $\varphi=0.5$ in most networks (except Physicians and Email), which means we need to consider the global or local differences between networks for comparison. Therefore, we use $\varphi=0.5$ in the above analysis.

\section{Conclusion}
\label{5}
In this paper, we introduce a comparison method $D_m$ that utilizes network motifs to assess similarities in directed networks. The method, which considers both local and global differences between two directed networks as well as higher-order information, is based on node motif distributions and employs Jensen-Shannon divergence. In detail, we use the motifs of sizes up to 4 that are listed in Figure ~\ref{fig:motif} to compute the motif distribution of nodes in a directed network. Based on Jensen-Shannon divergence and motif distributions of nodes, we define the dispersion of directed network nodes ($DNND$) to quantify the heterogeneity of connectivity between nodes. Lastly, for two given directed networks, the similarity between them is further defined by the combination of the $DNND$ metrics and the average motif distributions. Our method aims to better understand the internal connection patterns of the network nodes by capturing essential subgraph structures. To show the effectiveness of our method, we compare a directed network with its null models, which gradually change the structure of the original network. In addition, we further compare our method with the baselines to characterize the similarity between an orignal network and its perturbed networks
The results show that our method outperforms these baseline methods across networks from different domains.

Motifs have been widely used to address a range of tasks. In our analysis, we take into account the directionality of edges by utilizing directed motifs to compare directed networks. We limit our analysis to motifs of sizes up to 4 due to the high computational expenses involved. Although considering larger motifs could potentially enhance the effectiveness of our approach, it may pose scalability challenges when dealing with large networks containing millions of nodes. Given the success of motifs in network comparison, we believe that developing efficient algorithms for computing motifs could be a promising avenue for research. This not only has the potential to enhance network comparison, but also to improve other network tasks such as community detection, node classification, influence maximization, and more.

\section{Declaration of competing interest}
The authors declare that they have no known competing financial interests or personal relationships that could have appeared to influence the work reported in this paper.
\section{Data availability}
Data will be available on request.
\section{Acknowledgement}
This work was supported by the Natural Science Foundation of Zhejiang Province (Grant No. LQ22F030008), the Natural Science Foundation of China (Grant No. 61873080), and the Scientific Research Foundation for Scholars of HZNU (2021QDL030).

\bibliographystyle{elsarticle-num}
\bibliography{TempExample}

\end{document}